\documentclass[preprint,showpacs,preprintnumbers,amsmath,amssymb,floatfix]{revtex4}

\usepackage{bm}% bold math

\newcommand{\Vp}{{\bf p}}
\newcommand{\Vq}{{\bf q}}

\begin{document}

%\preprint{APS/123-QED}

\title{Issues on Neutrino-Nucleus Reactions \\ in the Quasi-free Delta
Production Region}

\author{Ryoichi Seki${}^{1,2}$ and Hiroki Nakamura${}^3$}

\affiliation{${}^1$ Department of Physics and Astronomy,
California State University, Northridge, Northridge, CA 91330 \\
${}^2$ W. K. Kellogg Radiation Laboratory, California Institute
of Technology, Pasadena, CA 91125 \\
${}^3$ Department of Physics, Waseda University, Tokyo 169-8555,
Japan}

\date{\today}

\begin{abstract}
A brief overview is presented of various issues involved in
phenomenological and theoretical works on charge-current
neutrino-nucleus reactions associated with the quasi-free
$\Delta(1232)$ production.  An assessment of the present status of
the works is made with respect to the objective of this
conference, Sub-Dominant Oscillation Effects in Atmospheric
Neutrino Experiments.
\end{abstract}

\maketitle

\section{Introduction}

Above the quasi-elastic neutrino-reaction energy, the prominent
process in the charge-current neutrino-nucleus reactions is the
quasi-free $\Delta(1232)$ production.  We present a brief overview
of the present phenomenological and theoretical studies on the
process and discuss what one could focus on for further progress.
Note that higher-mass resonance and multi-pion productions, as
well as a diffractive non-resonance background do contribute to
the neutrino reaction cross sections in this region.  They are
needed to be included for a realistic description of the neutrino
reactions in this region, and their treatment combined with the
$\Delta(1232)$ production requires more refined consideration.

This presentation is not meant to be a comprehensive review, but a
short overview based on the formalism described below, in which we
believe various physics involved in the process emerges most
clearly. We do not discuss, for example, work of Valencia/Granada
School, which follows a different formalism by systematically
applying a nuclear many-body theory starting from Fermi gas. The
reader can find their work on electron scattering in \cite{gno},
and on neutrino scattering in \cite{nav} and J. Nieves's talk in
this conference.

The inclusive neutrino-nucleus $(\nu,\ell)$ cross section is
written in linear response formalism as
\begin{equation}
\frac{d\sigma}{dE_\ell d\Omega_\ell}=
\frac{k_\ell}{8(2\pi)^4M_A^2E_\nu}
  \int d^3\Vp F({\bf p},{\bf q},\omega) \left|{\cal M}_{\nu N}\right|^2
 \;,
\end{equation}
where $E_\nu$ is the incident neutrino energy, $E_\ell$ and
$\Omega_\ell$ are the energy and solid angle of the scattered
lepton, respectively, and $M_A$ is the target nuclear mass.  $\cal
M_{\nu \rm{N}}$ is the invariant on-shell neutrino-nucleon
scattering amplitude, depending on the Mandelstam variables
(expressed in terms of the four momenta of the leptons and the
nucleon). $F(\Vp,\Vq,\omega)$ includes all relevant information of
the initial nuclear state and also the final-state interactions
between the lepton and the final-state nucleus. $F$ depends on
$\Vp$, the momentum of the initial nucleon, and $\Vq$ and
$\omega$, the momentum and energy transfer to the nucleus,
respectively.

$F(\Vp,\Vq,\omega)$ describes the response of the nucleus to the
disturbance generated by the neutrino probe, $\left|{\cal M}_{\nu
N}\right|^2$.  Here, we are applying the widely used approximation
that neutrino interacts only directly with the nucleon.  (See also
Subsection 3.1.) $F(\Vp,\Vq,\omega)$ is expressed in the form of
two-particle Green's function as
\begin{eqnarray}
F({\bf p},{\bf q},\omega) &=& \langle A |a^\dagger_{\bf p}a_{\bf
p+q} \delta(\hat{H}-E_A-\omega)a^\dagger_{\bf p+q}a_{\bf p}|A
\rangle \nonumber\\
&\approx& \frac{1}{2M_A} \int {d\omega'}
P_h({\Vp},\omega')P_p({\Vp}+{\Vq},\omega-\omega')\;,
\label{2green}
\end{eqnarray}
where $a_\Vp$ and $a^\dagger_\Vp$ are the annihilation and
creation operators of a nucleon, $|A \rangle$ is the target
nucleus in the ground state, and $E_A$ is its energy.  The delta
function (operator) ensures the final state of the process to be
physical (on-shell).  In the second step in Eq. (\ref{2green}),
$F(\Vp,\Vq,\omega)$ is approximately factorized as a product of
single-hole and single-particle Green's functions,
$P_h({\Vp},\omega)$ and $P_p({\Vp},\omega)$, respectively.
$P_h(\Vp,\omega)$ is (apart from a simple kinematic factor) the
probability of finding a nucleon of the momentum $\Vp$ and removal
energy $\omega$ in the nucleus, and is referred as the spectral
function. $P_p(\Vp,\omega)$ is a similar quantity for adding a
nucleon in the nucleus and contains information of the final-state
interactions.

$\left|{\cal M}_{\nu N}\right|^2$ is the $\nu N$ cross section
apart from a kinematic factor, and
\begin{equation}
\left|{\cal M}_{\nu N}\right|^2 \propto \eta_{\mu\nu}T^{\mu\nu}\;,
\end{equation}
where $\eta_{\mu\nu}$ is the leptonic tensor and $T^{\mu\nu}$ is
the nucleonic tensor.  For the quasi-elastic reaction,
$T^{\mu\nu}$ is expressed as a product of the nucleon current. For
the quasi-free $\Delta$ production, it is expressed as a product
of the $N-\Delta$ transition currents $J$'s:
\begin{equation}
T_{\mu\nu} = \sum_{spin} J_\mu^-J_\nu^+ \times ({\rm
Breit-Wigner})\;,
\end{equation}
where (Breit-Wigner) describes the decaying state of
$\Delta$(1232). $J$'s expressed as a linear combinations of the
transition form factors $C$'s
\begin{equation}
J_\mu^+ =
\bar{U}^\rho(p')\Gamma_{\rho\mu}(C^V_3(-t),C^V_4(-t),C^V_5(-t),C^A_5(-t);p',p)
u(p)\;,
\end{equation}
where $U^{\rho}$ and $u$ are the spin 3/2 (Rarita-Schwinger)
spinor and the nucleon spinor, respectively.   The invariant
principles dictates that $\Gamma$ depends also on other form
factors, but $C^V_3(-t)$ and $C^A_5(-t)$ are most dominant and
closely studied.

We will first examine in Section 2 lepton-nucleon reactions,
because they serve as an input for nuclear reaction calculations;
and second, nuclear reactions in Section 3. The concluding remarks
are presented in Section 4.

\section{Input: Lepton-Nucleon Reactions}

\subsection{Electron scattering}

Around 1980, Bodek et al.\cite{b79,b81} analyzed inclusive
$(e,e')$ SLAC data from $p$ and $D$, and extracted $p$ and $n$
responses, $W_1$ and $W_2$, phenomenologically.  $W_1$ and $W_2$
are related to the inclusive cross section through the well-known
expression,
\begin{equation}
\frac{d^2\sigma}{d\Omega dE_\ell} =
\sigma_{Mott}[W_2+2\tan^2\theta \cdot W_1]\;,
\end{equation}
where $\sigma_{Mott}$ is the Mott scattering cross section. As
this expression shows, $W_1$ and $W_2$ are inclusive quantities,
including all other processes in addition to the $\Delta(1232)$
production.

Subsequently, exclusive, detailed data have been obtained at
Jefferson Lab, and their analyses have been carried out by the use
of phenomenological/theoretical models. The data and analyses have
been recently reviewed by Burkert and Lee \cite{bl04}, which
serves as a useful reference.

As described in the review, several theoretical models
(approaches) have been considered:

\begin{itemize}
\item Unitary isobar model, MAID \cite{maid} and Jlab/Yeveran
\cite{yev}.
\item Multi-channel K-matrix model, SAID \cite{said}.
\item Dynamical model, Sato-Lee \cite{sl} and Dubna-Mainz-Taiwan
\cite{dmt}.
\end{itemize}

All these models seem to have an excellent agreement with proton
data of explicit single pion production observables in the region
relevant to the $\Delta(1232)$ production. In addition to these
theoretical models, we add a phenomenological model,
\begin{itemize}
\item H2/D2 model \cite{h2}.
\end{itemize}
The H2 model was constructed using a large body of inclusive
electron-proton data from SLAC \cite{ck}, but was found to agree
with new data from Jefferson Lab \cite{IN1,YL} to better than 5 \%
as well. The D2 model is a fit to Jefferson Lab data \cite{IN2}
only.

Neutron $(e,e')$ data are scarce and with larger uncertainties, as
they have to be extracted from $D(e,e'p)$.  New Jefferson Lab data
are found in \cite{phd}.  A new precision experiment BONUS
(E03-012) is being planned at Jefferson Lab by measuring the
recoil proton [Private communication with C. E. Keppel.]  We note
that new precision experiments are also being carried out at
Jefferson Lab on deuterium [E02-109; Spokespersons, M. E. Christy
and C. E. Keppel] and also on nuclei [E04-001 (Jupiter);
Spokespersons, A. Bodek and C. E. Keppel.]

Overall, precise electron reaction data and their reliable
analyses are (and are becoming) available, so as to determine the
vector transition form factors rather well.

\subsection{Neutrino Reactions}

Following the detailed theoretical work of Adler \cite{adler} and
the comprehensive review of Llewellyn Smith \cite{lsmith} around
1970, Rein-Sehgal analysis \cite{rein} in 1980's has been serving
as the standard description of the resonance production processes.
The single pion production process has been re-examined in the
last several years \cite{asv, sul, pas}. The $N-\Delta$ transition
form factors used in these works are similar to each other, but
are different in detail.  Single-pion production cross sections by
the use of the form factors are usually compared to ANL \cite{anl}
and BNL \cite{bnl} data from the proton and deuterium targets.  As
the data are weighted with the neutrino flux energy distribution,
they serve only as weak constraints on the form factors.

The different form factors that have been used do affect nuclear
calculations and yield different cross sections, sometimes
appreciably.  The data were reported two decades ago. Clearly new
measurements are critically needed by the use of improved
technology, so as to determine the axial transition form factors
reliably.

\section{Neutrino-Nucleus Reactions}

Neutrino-nucleus reactions based on the present formalism are
discussed by H. Nakamura in this conference, focusing on
quasi-elastic scattering by incorporating the inclusive
electron-nucleus scattering data.  As most discussions are also
applicable to quasi-free $\Delta$ production, we will not repeat
them here.  Note that H. Nakamura's talk also includes some
results in quasi-free $\Delta$ production.

Instead, here we discuss what we believe to be prominent issues in
the treatments of nuclear structure and reactions in neutrino
process. The issues are:

%\vfill\eject

\begin{itemize}
\item Spectral functions.
\item Final-state interactions.
\item Exchange current.
\item Other approaches.
\end{itemize}

\subsection{Spectral Functions}

As noted in Section 1, the initial nuclear state needed is the
one-hole state of the target nucleus, as the hole state is
generated by the neutrino knocking out a nucleon.  The spectral
function describes this state as a function of energy to remove a
nucleon and of its momentum.  The energy is not discrete because a
hole state has a finite life time.  Deeply bound states such as
those of the s-shell have a quite broad distribution.  The
momentum distribution is also spread out, extending beyond the
Fermi momentum and beyond the momenta involved in shell-model and
mean-field calculations.  This is because of short-range nucleon
correlations of high-momentum components, generated by the
short-distance nuclear interactions. These features are much
different from those of simple Fermi gas model, which continues to
be widely used in the Monte Carlo analysis of experimental data.
Note that high-momentum nucleons tend to alter, for example, the
angular distributions of neutrino-nucleus reactions, though they
affect less the total cross sections.

When the removal energy is summed, the spectral function yields
the momentum distribution of a nucleon in the nucleus. The high
momentum components play an important role in neutrino-nucleus
reactions in the GeV region. A useful general review of nuclear
momentum distributions is found in \cite{ahp}.

The most detailed calculation of the spectral function has been
carried out by O. Benhar and his collaborators for ${}^{12}$C,
${}^{16}$O, and other nuclei \cite{benhar}, based on a nuclear
many-body calculation with correlated nuclear-state basis,
combined with shell model and local density approximation. Cross
sections calculated by the use of these spectral functions have
been reported at the NuInt04 conference on neutrino scattering in
comparison to the case of electron scattering \cite{nuint04}. Note
that different recent calculations of the spectral functions are
available, for example, by Ciofi degli Atti and his collaborators
based on a simpler but more physical approach \cite{ciofi}.

\subsection{Final State Interactions}

For the outgoing nucleon, eikonal approximation has been applied
within an optical potential description \cite{ben} to inclusive
neutrino-nucleus reactions \cite{final}. The present application
remains to be of a simple estimate, and should be improved.

Monte Carlo simulation codes used in data analysis are basically a
classical description, and they, as well as the eikonal
calculations, need to include nuclear medium corrections in
nucleon-nucleon cross sections. The corrections are quite
substantial in low-energy nucleon-nucleon scattering \cite{pp},
which is involved in the final-state interactions of low-momentum
transfer (to the nucleon).

Let us note here on the occasionally raised question of how to
treat the final-state interactions by differentiating the
inclusive process (no outgoing nucleon is measured) and the
semi-inclusive process (the outgoing nucleon is measured): The
difference in two theoretical treatments is clearly understood in
the optical potential description \cite{hori}.

For the outgoing pion, a Monte Carlo algorithm, originally
developed for pion physics \cite{sal}, is currently in use in the
Monte Carlo simulation codes and is under a good control. Note,
however, that a low-energy pion in nuclear medium receives a
strong dispersive effect (described by the real part of the
pion-nucleus optical potential) in addition to an absorptive
(collision) effect (described by the imaginary part), as also
noted in \cite{sal}. The former is significant for low-energy
pions, for which the cross sections are strongly energy-dependent,
and should be included in Monte Carlo simulation.

\subsection{Exchange Current}

Almost all high-energy neutrino-nucleus calculations and Monte
Carlo simulation codes do not account for effects associated with
exchange currents in nuclei.  Thus, for example, off-mass/energy
shell contributions are not included in them, and theoretically
most importantly, the current conservation is violated.  Proper
inclusion of the effects is difficult, and is not done
satisfactorily even in electron scattering works.

Physics of this issue may play, however, a more important role in
high energy neutrino reactions than in corresponding electron
reactions, because of the axial currents. Quenching of 20 - 30 \%
in $g_A (GT)$ is well known in beta decays, and such effect has
been incorporated in some of solar neutrino calculations.  In
high-energy neutrino reactions, we may have a strong modification
of the form factors themselves beyond that of the coupling
constants.  This issue has been studied previously \cite{so}, but
would deserve a closer examination.  It is a complicated matter,
as pion dynamics in nuclei has to be carefully sorted out,
together with the parallel view of a possible modification of
nucleon structure in nuclei. Note that a series of the many-body
theory investigation of the axial form factors in nuclei has been
made by D. Riska and his collaborators \cite{riska}.

In this connection, we emphasize that the rigorous determination
is vital of the axial form factors of the nucleon in free space.
There is an urgent need for more detailed, reliable neutrino
scattering experiments from the proton and deuterium. Furthermore,
the exchange effects in neutrino-deuterium reactions perhaps
should be re-examined for the extraction of the axial form
factors, as the existing work is nearly two decade old \cite{sa}.

\subsection{Other Approaches}

\begin{itemize}
\item We have already noted work of Valencia/Granada School, which
uses a different many-body theory approach, incorporating many of
comments made here.  Please see J. Nieves's talk in this
conference.

\item T. W. Donnelly , I. Sick, and their collaborators proposed
a (super)scaling approach \cite{sca1} based on the observation
that the simple Fermi gas model works fairly well in electron
scattering.  An extension to neutrino scattering has been recently
reported \cite{sca2}.  In this approach, it has not been clarified
what physics is included in the modifying factor of the Fermi gas
model, and thus how reliable the extension to neutrino scattering
is from electron scattering, because the currents involved in the
two processes are different.

\end{itemize}

\section{Summary}

We conclude this presentation by briefly assessing the present
status of works in the Delta region by listing what we consider to
be its most important aspects:

\begin{itemize}
\item Quality:  Presently available calculations and codes
include most of important physics at various degrees, with the
exception of the physics associated with the current conservation.
\item Predictability:  More tuning of the calculations is
desirable to precise electron-scattering data currently available
or becoming available.
\item Experiment: More precise neutrino-nucleon data are
critically needed.
\item Codes: It is perhaps the best time for upgrading the
existing codes to create codes of the next generation by including
all available nuclear-physics information and knowledge that have
been accumulated, for detailed examination of atmospheric neutrino
experiments and for the upcoming high quality experiments such as
those at J-Parc and Fermi Lab.

\end{itemize}

%\vfill\eject
\bigskip
\bigskip
\bigskip

{\bf Acknowledgment}

\bigskip

We acknowledge C. E. Keppel and M. E. Christy for providing us the
update information of electron scattering, and M. Sakuda for
continuing collaboration on high-energy neutrino-nucleus
reactions.  This work is supported by the U. S. Department of
Energy under grant DE-FG03-87ER40347 at CSUN and by the U. S.
National Science Foundation under grant 0244899 at Caltech.


\bigskip
\bigskip
\bigskip


\begin{thebibliography}{99}%%%%%%%%%%

\bibitem{gno} A. Gil, J. Nieves, and E. Oset, Nucl. Phys. A627
(1997) 543; and references therein.
\bibitem{nav} J. Nieves, J. E. Amaro, and M. Valverde,
nucl-th/0408008.
\bibitem{b79} A. Bodek et al., Phys. Rev. D20 (1979) 1471.
\bibitem{b81} A. Bodek and J. L. Richie, Phys. Rev. D23 (1981),
1070.
\bibitem{bl04} V. D. Burkert and T.-S. H. Lee, nucl-ex/0407020.
\bibitem{maid} D. Drechsel, S. S. Kamalov, and L. Tiator, Nucl.
Phys. A645 (1999) 145.
\bibitem{yev} I. G. Aznauryan, Phys. Rev. C67 (2003) 015209; ibid.,
C68 (2003) 065204.
\bibitem{said} R. A. Arndt et al., Phys. Rev. C52 (1995) 2120;
Int. J. Mod. Phys. A18 (2003) 449.
\bibitem{sl} T. Sato and T.-S. H. Lee, Phys. Rev. C54 (1996)
2660; ibid., C63 (2001) 055201.
\bibitem{dmt} S. S. Kamalov and S. N. Young, Phys. Rev. Lett. 83
(1999) 4494; and D. Drechsel, O. Hanstein, and L. Tiator, Phys.
Rev. C64 (2002) 032201.
\bibitem{h2} C. E. Keppel, http://hallcweb.jlab.org/resdata/ and
private communication with M. E. Christy.
\bibitem{ck} C. Keppel, PhD thesis (American Univ.), SLAC-R-694
(1994).
\bibitem{IN1} I. Niculescu et al., Phys. Rev. Lett. 85 (2000) 1186.
\bibitem{YL} Jefferson Lab Hall C E94-110 Collaboration,
nucl-ex/0410027.
\bibitem{IN2} I. Niculescu et al., Phys. Rev. Lett. 85 (2000) 1182.
\bibitem{phd} A. V. Klimenko, PhD thesis (Old Dominion Univ.,
2004).
%\bibitem{bonus) Private communication with C. Keppel.
%\bibitem{sakuda} Private communication with M. Sakuda.
%\bibitem{jup} Spokesperson, C. Keppel and A. Bodek.
\bibitem{adler} S. L. Adler, Ann. Phys. (N. Y.) 50 (1968) 189.
\bibitem{lsmith} C. H. Llewellyn-Smith, Phys. Rept. 3 (1972) 261.
\bibitem{rein} D. Rein and L. Sehgal, Ann. Phys. 133 (1981) 79;
and D. Rein, Z. Phys. C35 (1987) 43.
\bibitem{asv} L. Alverez-Ruso, S. K. Singh, and M. J. Vicente
Vacas, Phys. Rev. D69 (2004) 014013.
\bibitem{sul} T. Saito, D. Uno, and T.-S. H. Lee, Phys. Rev. C67
(2003) 065201.
\bibitem{pas} E. A. Paschos, L. Pasquali, and J.-Y. Yu, Nucl.
Phys. B588 (2000) 263; and E. A. Paschos, J.-Y. Yu, and M. Sakuda,
Phys. Rev. D69 (2004) 014013.
\bibitem{anl} S. J. Barish et al., Phys. Rev. D19 (1979) 2521.
\bibitem{bnl} T. Kitagaki et al., Phys. Rev. D34 (1986) 2554.
\bibitem{ahp} A. N. Antonov, P. E. Hodgson, and I. Zh. Petkov,
{\it Nucleon Momentum and Density Distributions in Nuclei}
(Clarendon Press, Oxford, 1988).
\bibitem{benhar} Private communication with O. Benhar;
O. Benhar, A. Fabrocini, S. Fantoni, and I. Sick, Nucl. Phys. A579
(1994) 493; and references therein.
\bibitem{nuint04} O. Benhar (in {\it Proc.
Third Int. Workshop on Neutrino-Nucleus Int. in the few GeV
region}), Nucl. Phys. B.--Proc. Suppl. 139 (2005) 15; H. Nakamura,
R. Seki and M. Sakuda, ibid. 139 (2005) 201; and O. Benhar and N.
Farina, ibid. 139 (2005) 230.
\bibitem{ciofi} C. Ciofi degli Atti, D. B. Day, and S. Liuti,
Phys. Rev. C46 (1992) 1045.
\bibitem{ben} O. Benhar et al., Phys. Rev. C44 (1991) 2328.
\bibitem{final} H. Nakamura, this proceedings; and O. Benhar,
private communication.
\bibitem{pp} V. R. Pandharipande and S. C. Pieper, Phys. Rev. C45
(1992) 791.
\bibitem{hori} Y. Horikawa, F. Lenz, and N. C. Mukhopadhyay, Phys.
Rev. C22 (1980) 1680.
\bibitem{sal} L. L. Saalcedo, E. Oset, M. J. Vicente Vacas, C.
Garcia Recio, Nucl. Phys. A484 (1988) 557.
\bibitem{so} S. K. Singh and E. Oset, Nucl. Phys. A5422 (1992)
587.
\bibitem{riska} M. Kirchbach and D. O. Riska, Nucl. Phys. A578
(1994) 511; M. Hjorth-Jensen et al., Nucl. Phys. A563 (1993) 525;
and D. O. Riska, Phys. Rep. 181 (1989) 208.
\bibitem{sa} S. K. Singh and H. Arenh\"{o}vel, Z. Phys. A324
(1986) 347.
\bibitem{sca1} C. Marieron, T. W. Donnelly, and I. Sick, Phys.
Rev. C65 (2002) 025502.
\bibitem{sca2} J. E. Amaro et al., nucl-th/0409078.

\end{thebibliography}
\end{document}